# Experimental demonstration of a broadband array of invisibility cloaks in the visible frequency range

V. N. Smolyaninova [1], I. I. Smolyaninov [2], H. K. Ermer [1]

[1] *Department of Physics Astronomy and Geosciences, Towson University,*

*8000 York Rd., Towson, MD 21252, USA*

[2] *Department of Electrical and Computer Engineering, University of Maryland,*

*College Park, MD 20742, USA*

**Very recently Farhat *et al.* [1] have suggested that arrays of invisibility cloaks may find important applications in low-interference communication, noninvasive probing, sensing and communication networks, etc. We report on the first experimental realization of such an array of broadband invisibility cloaks, which operates in the visible frequency range. Wavelength and angular dependencies of the cloak array performance have been studied.**

Ever since the first experimental demonstrations in the microwave [2] and visible [3] ranges, invisibility cloaks paved the way to considerable progress in the fields of metamaterials and transformation optics. Based on the considerable recent progress in the cloaking research, very recently Farhat et al. [1] introduced a new concept of an invisibility cloak array, which may find numerous applications in such fields as low-interference communication, noninvasive probing, sensing and communication networks, etc. Besides these obvious practical applications, building and studying the arrays of invisibility cloaks offer more refined experimental tools to test cloak performance. Here we report on the first experimental realization of such an array of broadband invisibility cloaks, which operates in the visible frequency range.

Our experimental geometry is based on the 2D broadband invisibility cloak design which utilizes an adiabatically tapered gold-coated waveguide emulating anisotropic dielectric permittivity and magnetic permeability distributions required for realization of the transformation optics-based invisibility cloak [4] (see Fig.1(a)). This approach leads to low-loss, broadband performance in the visible frequency range, which is difficult to achieve by other means. This simple design has been extended to 3D cylindrical geometry and its broadband performance was independently verified in 3D [5]. The basic idea of this design may be easily understood in the ray-optics approximation based on the semi-classical 2D cloaking Hamiltonian (dispersion law) introduced in [6]:

$$\frac{\omega^2}{c^2} = k_r^2 + \frac{k_\phi^2}{(r-b)^2} = k_r^2 + \frac{k_\phi^2}{r^2} + k_\phi^2 \frac{b(2r-b)}{(r-b)^2 r^2} \qquad (1)$$

Jacob and Narimanov demonstrated that for such a cylindrically symmetric Hamiltonian, the rays of light would flow smoothly without scattering around a cylindrical cloaked region of radius $b$. Such a cloaking Hamiltonian may be emulated by a gold-coated tapered waveguide, which thickness $d$ in the $z$-direction changes adiabatically with radius $r$. The dispersion law (Hamiltonian) of light in such a waveguide is

$$\frac{\omega^2}{c^2} = k_r^2 + \frac{k_\phi^2}{r^2} + \frac{\pi^2 m^2}{d(r)^2} \ , \qquad (2)$$

where $m$ is the transverse mode number. It appears that the cloaking Hamiltonian (1) can be emulated by an adiabatically changing $d(r)$. Moreover, the required waveguide shape is very close to a gap between a sphere and a plane surface [4]. The cloak radius for a mode number $m$ is then given as

$$r_m = \sqrt{(m+1/2)R\lambda} \,, \qquad\qquad (3)$$

where $R$ is the sphere radius and $\lambda$ is the wavelength of light. This cloaking geometry appears to be broadband with the cloaked areas for different light wavelengths nested inside each other [4]. As demonstrated in Fig.1(a), the described geometry is easy do transform into an array of broadband invisibility cloaks using commercially available microlens arrays. In the ideal case scenario light would propagate through such an array without scattering, as demonstrated in Fig.1(b).

Our invisibility cloak arrays were fabricated as follows. As a first fabrication step, a commercially available microlens array [7] was coated on the microlens side with a 30-nm gold film (Fig.2a). The array was placed with the gold-coated side down on top of a flat glass slide coated with a 30-nm gold film. Two gold coated surfaces were pressed against each other using a mechanical arrangement with set screws. Argon ion laser light with different wavelengths $\lambda$ was coupled into the waveguide from the side. Periodic array of adiabatically tapered gaps between the gold-coated surfaces has been used as a 2D array of invisibility cloaks similar to the ones described in refs.[4,5].

As a first experimental step, we have studied 514 nm light propagation through a cloak array formed by an array of large microlenses (500 μm pitch, 56 mm lens radius) as shown in Fig.2. As clearly visible in Fig.2(b), laser beam diameter in this experiment is comparable with the distance between the individual cloaks in the array. Dashed line in Fig.2(b) indicates the waveguide edge (the top and bottom gold-coated surfaces did not overlap precisely), while light propagation direction is indicated by the arrow. Similar to ref.[4], cloaked areas appear as dark circles surrounded by concentric rings in the experimental images. According to eq.(3), roughly 20% of the surface area is cloaked in this experiment, which is consistent with the experimental image in Fig.2(c). While such images clearly demonstrate that an array of cloaks have been created in the experiment, cloak separation appears to be too large to study individual cloak interaction.

In order to clearly demonstrate the effects of cloak interaction, we have used smaller array parameters (30 µm pitch, 42 µm lens radius) in the next set of experiments. These results are presented in Figs. 3-5. From the basic symmetry considerations the hexagonal dense cloak array shown in these figures is supposed to work best while illuminated along its three main symmetry axis, as shown in Fig.3. Indeed, microscope image of 514 nm light propagation through the cloak array is consistent with the "idealized" cloaking behaviour as shown in the inset. Similar to ref.[4], cloaked areas appear as dark circles. The inset in Fig.3(b) illustrates light propagation through the array. We were also able to confirm broadband cloaking behaviour of the cloak array illuminated along one of the main symmetry axis. Images of the array taken using 514 nm (Fig.4(a)) and 488 nm (Fig.4(b)) laser light coupled into the waveguide from the side illustrate similar cloaking performance.

On the other hand, angular performance of the dense hexagonal cloak array clearly shows signs of deterioration. As illustrated in Fig. 5(d), cloak array illumination along the direction which does not coincide with one of the three main symmetry axis of the array leads to reduction of symmetry of the problem. This must lead to enhanced light scattering inside the array. Comparison of images in Fig.5(a) and Fig.5(c) indeed demonstrates such an enhanced light scattering inside the rotated cloak array.

In conclusion, we have reported the first experimental realization of an array of broadband invisibility cloaks, which operates in the visible frequency range. Such an array is able of cloaking ~20% of an unlimited surface area. We have studied wavelength and angular dependencies of the cloak array performance. While the broadband performance appears to be similar to the performance of individual cloaks in the array, angular performance of a dense array shows sign of deterioration due to reduction of symmetry of the cloaking arrangement.

**Acknowledgements**

This research was supported by the NSF grants DMR-0348939 and DMR-1104676. We are grateful to A. Piazza and R. Kuta for experimental help.

**Figure Captions**

**Figure 1.** (a) Experimental geometry of the broadband array of invisibility cloaks. A single gold-coated lens from ref.[3] has been replaced with a gold-coated microlens array. (b) Illustration of light propagation through an array of invisibility cloaks. The cloaked areas are shown in black.

**Figure 2**. Light propagation through a rectangular cloak array formed by the gap between gold coated surfaces of a large microlens array (500 µm pitch, 56 mm lens radius) and a flat glass slide: (a) Microscope image of the gold coated lens array. (b) Microscope image of 514 nm light propagation through the cloak array. Dashed line indicates the waveguide edge. Light propagation direction is indicated by the arrow. (c) Magnified image of 514 nm light propagation through the cloak array. Similar to ref.[4], cloaked areas appear as dark circles surrounded by concentric rings.

**Figure 3**. Light propagation through a hexagonal cloak array formed by the gap between gold coated surfaces of a microlens array (30 µm pitch, 42 µm lens radius) and a flat glass slide: (a) Microscope image of the gold coated lens array. (b) Microscope image of 514 nm light propagation through the cloak array. Light propagation direction is indicated by the arrow. Similar to ref.[4], cloaked areas appear as dark circles. Inset illustrates light propagation through the array. Three main symmetry axis of the hexagonal structure are shown by dashed lines.

**Figure 4**. Broadband performance of the cloak array is illustrated by two images of light propagation through the array taken using 514 nm (a) and 488 nm (b) laser light coupled into the waveguide from the side. Image (c) taken using combination of illumination with white light from the top and 514 nm laser light from the side further

illustrates that laser light travels around the central Newton ring areas of the cloaks, which appear bright under the white light illumination from the top.

**Figure 5**. Light propagation through the rotated cloak array indicates reduction in cloak array performance. Image (a) was taken at 514 nm when the cloak array was rotated by ~90$^{o}$. Orientation of the cloak array is demonstrated in image (b) taken under combination of white light illumination from the top and 514 nm laser light from the side. Comparison of images (a) and (c) clearly demonstrates enhanced scattering inside the rotated cloak array. This result may be explained by reduced symmetry of the problem, as illustrated in (d) where direction of light propagation does not coincide with one of the three main symmetry axis of the hexagonal array.

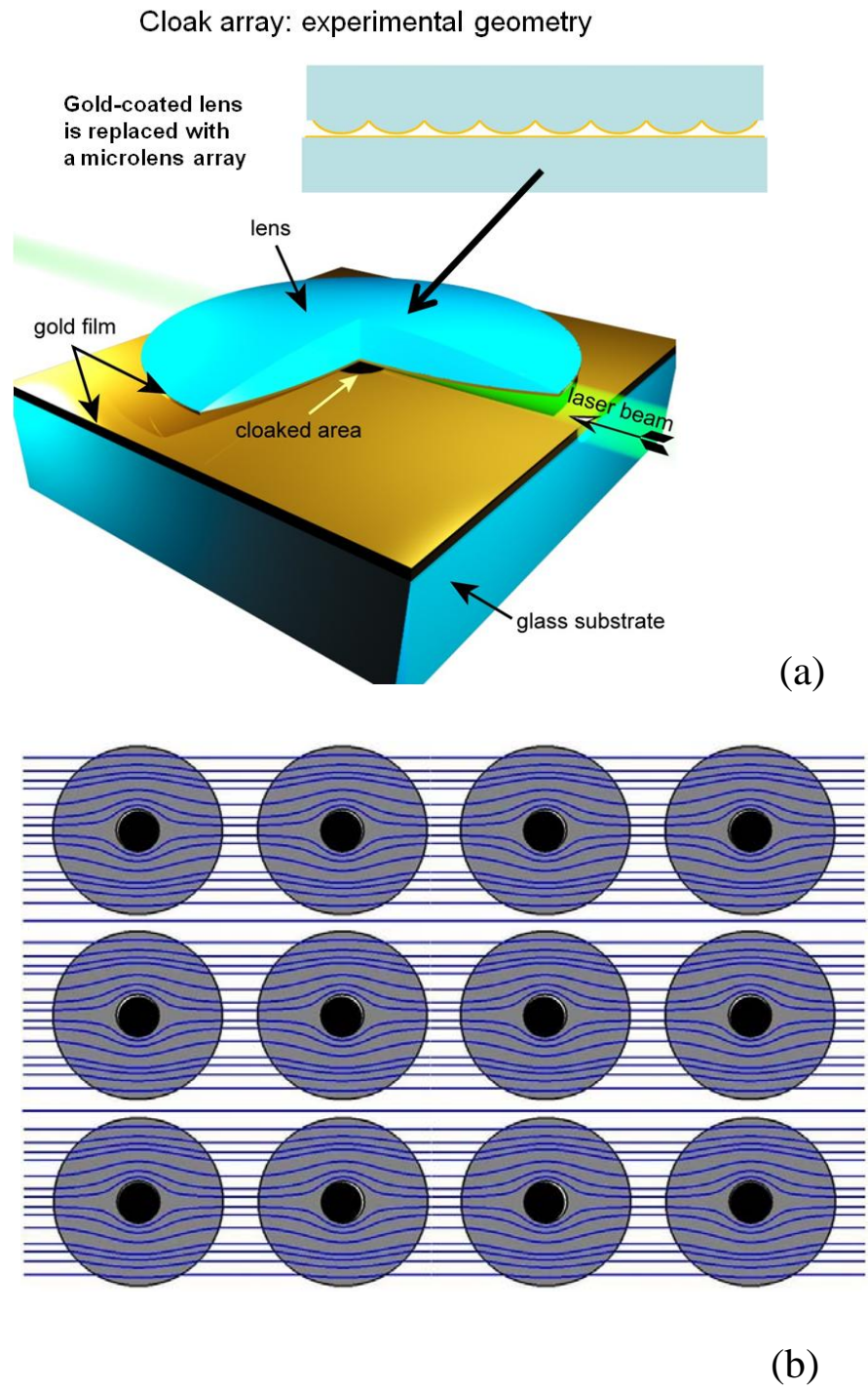

Cloak array: experimental geometry

Gold-coated lens is replaced with a microlens array

lens

gold film

cloaked area

laser beam

glass substrate

(a)

(b)

Fig.1

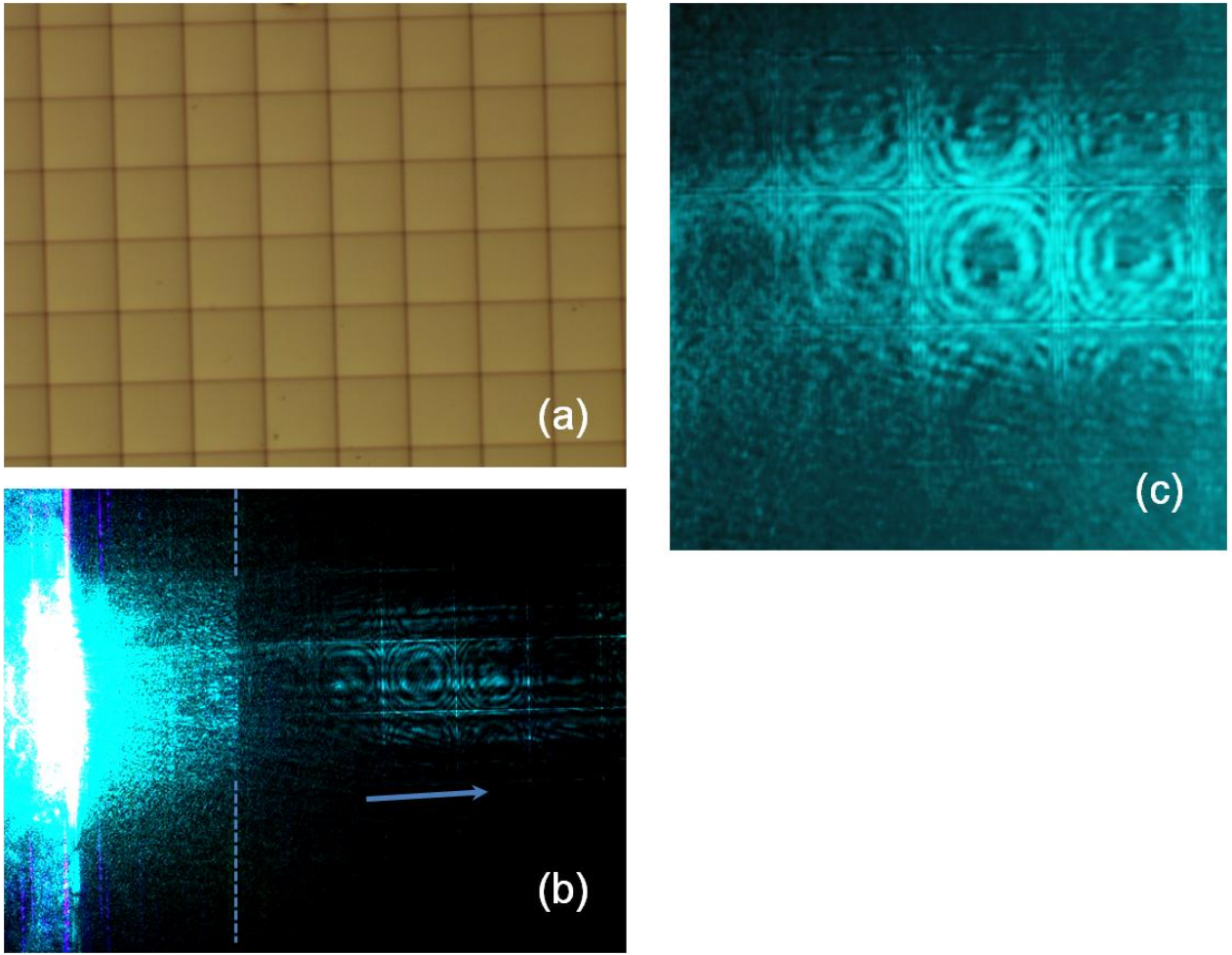

Fig. 2

(a)

(b)

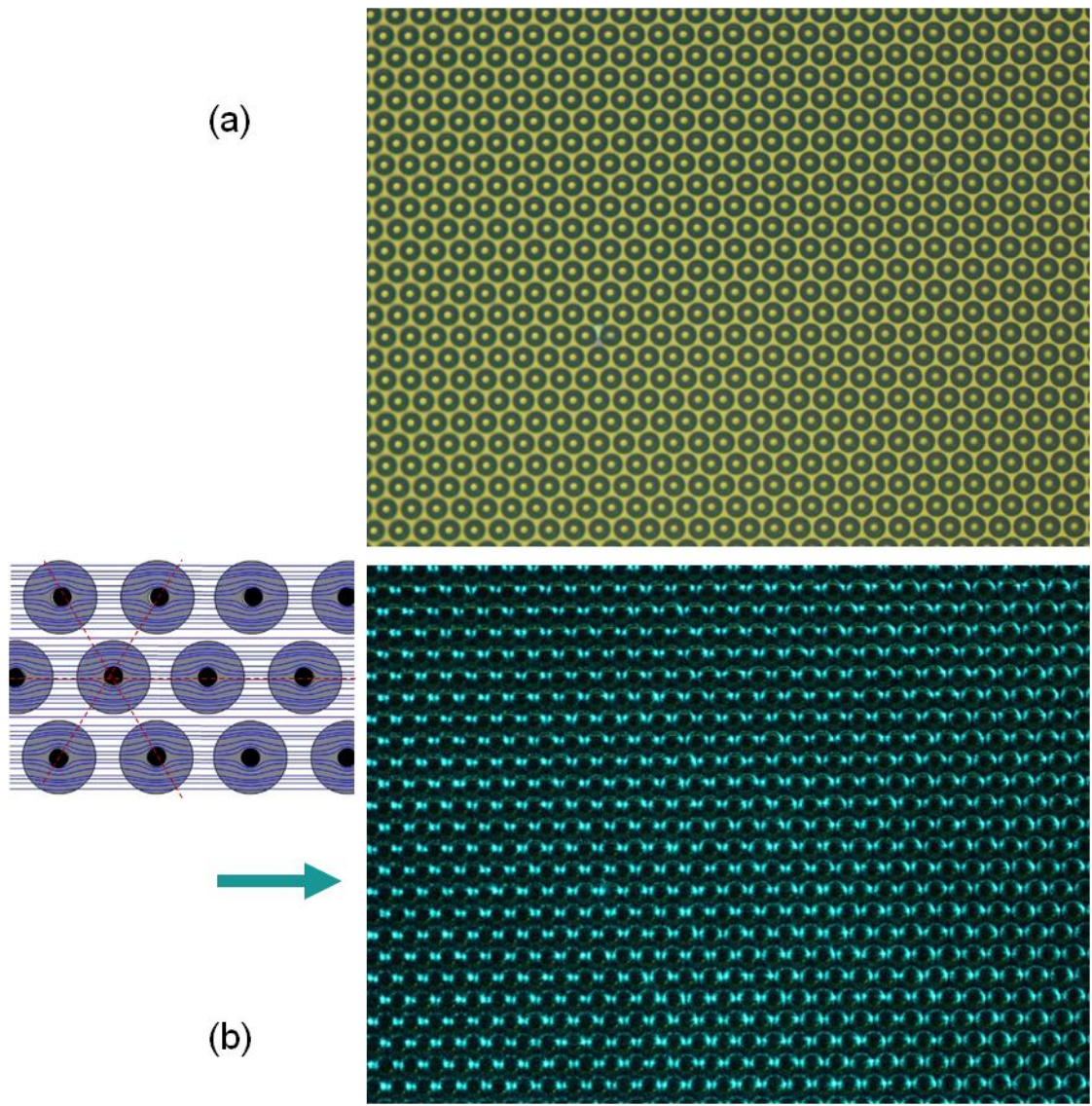

Fig. 3

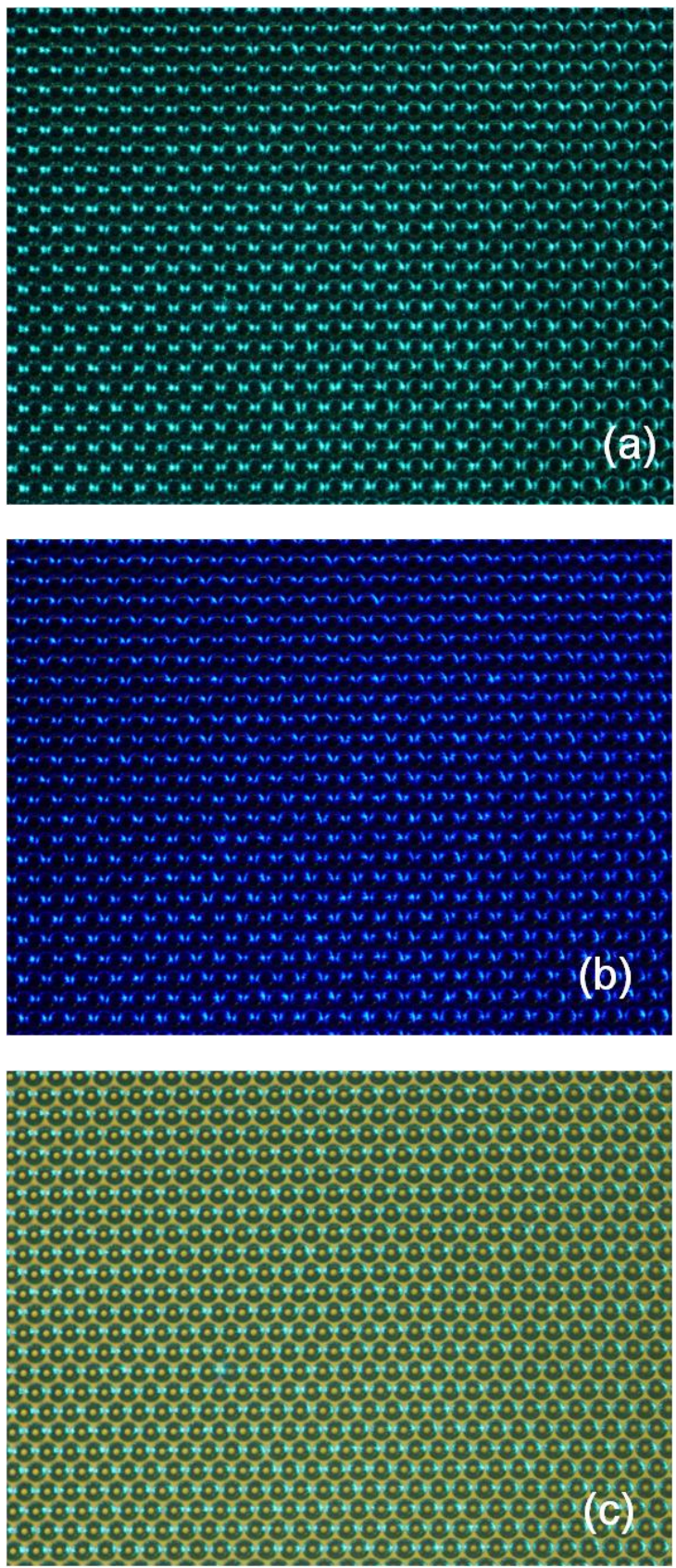

Fig. 4

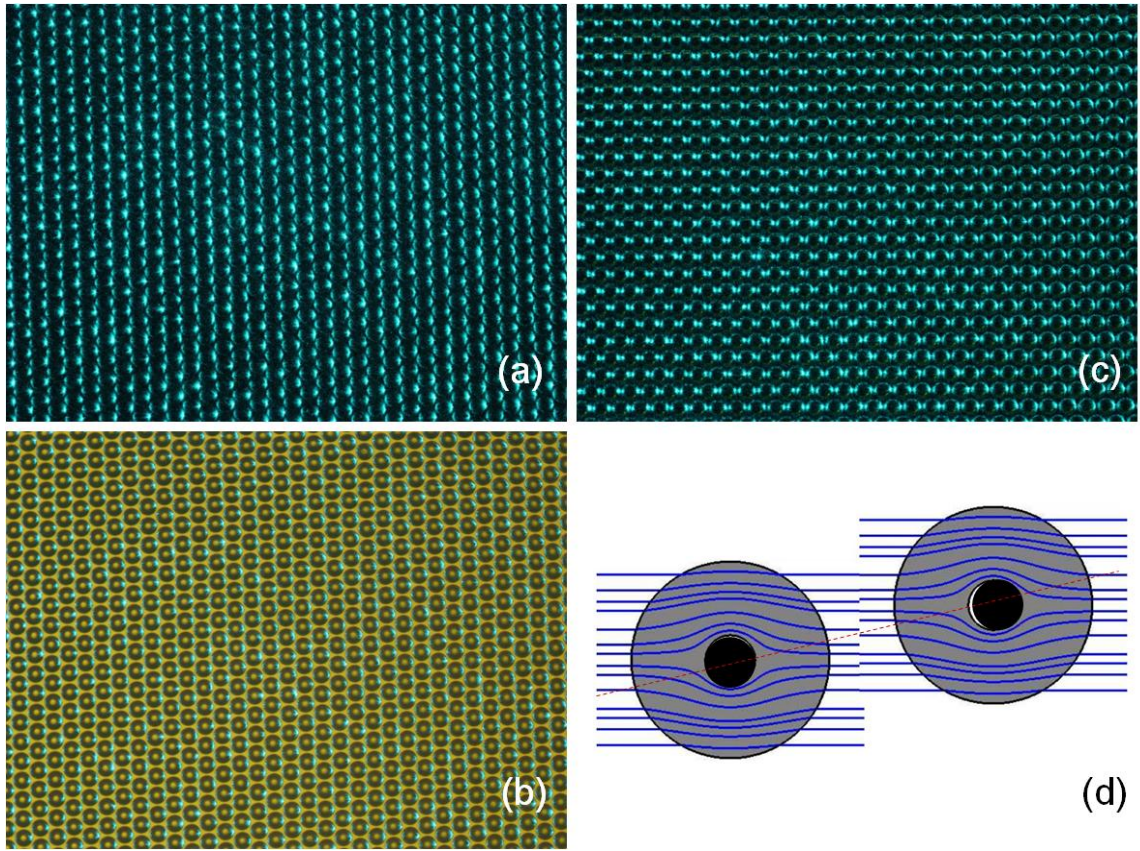

Fig. 5